\documentclass{aastex}

\usepackage{amsmath}
\usepackage{graphicx}

\usepackage{emulateapj5}
\def\muG{$\mu\,$G}
\newcommand{\degree}{\degr}

\makeatletter

\newenvironment{inlinefigure}{%
\def\@captype{figure}%
\noindent\begin{minipage}{0.999\linewidth}\begin{center}}
{\end{center}\end{minipage}\smallskip}
\makeatother

\def\slantfrac#1#2{\mbox{\raisebox{.4ex}{$\scriptstyle #1$}\kern-0.1em$/$\kern-0.1em\raisebox{-.4ex}{$\scriptstyle #2$}}}
\def\Im{{\rm Im}}

\def\inf{_{\scriptscriptstyle \infty}}

\usepackage{timesfonts}

\def\keV{ke\kern-0.05emV}

\newcommand{\chandraintitle}{{\itshape\footnotesize CHANDRA}}
\newcommand{\KH}{K-H}

\begin{document}

\title{\emph{Chandra} estimate of the magnetic field strength near the cold
  front in A3667}


%
%
%

\author{A.\ Vikhlinin\altaffilmark{1}, M.\ Markevitch, S.\ S.\ Murray}
\affil{Harvard-Smithsonian Center for Astrophysics, 60 Garden St.,
Cambridge, MA 02138;\\ avikhlinin@cfa.harvard.edu}

\altaffiltext{1}{Also Space Research Institute, Moscow, Russia}

\shorttitle{MAGNETIC FIELD IN A3667}
\shortauthors{VIKHLININ, MARKEVITCH, \& MURRAY}

\begin{abstract}
  
  We use the \emph{Chandra} observation of the cold front in the
  intracluster gas of A3667 to estimate the magnetic field strength near the
  front. The front is seen in the \emph{Chandra} data as a sharp
  discontinuity in the gas density which delineates a large body of dense
  cool gas moving with the near-sonic velocity through the less dense,
  hotter gas. Without a magnetic field, the front should be quickly
  disturbed by the Kelvin-Helmholtz instability arising from tangential
  motion of gas layers.  However, \emph{Chandra} image shows that the front
  is stable within a $\pm30\degree$ sector in the direction of the cloud
  motion, beyond which it gradually disappears. We suggest that the
  Kelvin-Helmholtz instability within the $\pm30\degree$ sector is suppressed
  by surface tension of the magnetic field whose field lines are parallel to
  the front.  The required field strength is $B\sim10\,$\muG. Magnetic field
  near the front is expected to be stronger and have very different
  structure compared to the bulk of the intergalactic medium, because the
  field lines are stretched by the tangential gas motions. Such a magnetic
  configuration, once formed, would effectively stop the plasma diffusion
  and heat conduction across the front, and may inhibit gas mixing during
  the subcluster merger. We note that even the increased magnetic field near
  the front contributes only 10--20\% to the total gas pressure, and
  therefore magnetic pressure is unimportant for hydrostatic cluster mass
  estimates.
  

\end{abstract}


\section{Introduction}

Magnetic fields can profoundly affect the properties of the intergalactic
medium in clusters. Examples include suppression of heat conduction, which
is required to maintain the cluster cooling flows (see the review by Fabian
1994), and a possibility of a dynamically significant magnetic pressure to
explain the difference between the X-ray and strong lensing cluster mass
estimates (Loeb \& Mao 1994, Miralda-Escud\'e \& Babul 1995). So far,
intracluster magnetic fields were measured by Faraday rotation in radio
sources seen through the IGM (e.g., Kim, Kronberg \& Tribble 1991), or by
combining radio and hard X-ray data on cluster radio halos under the
assumption that the X-rays ares produced by inverse Compton scattering of
the microwave background (e.g., Fusco-Femiano et al.\ 2000).  Both these
methods indicate the magnetic field strength on the microgauss level, with
considerable uncertainty. In this Paper, we apply a completely new approach
that uses the effect of the magnetic field on the dynamics of the
intracluster plasma.

\emph{Chandra} recently revealed that in at least two clusters,A2142
(Markevitch et al.\ 2000) and A3667 (Vikhlinin, Markevitch \& Murray 2000,
Paper I hereafter), the dense cool cores are moving with high velocity
through the hotter, less dense surrounding gas. In both cases, the cool gas
is separated from the hotter ambient gas by a sharp density discontinuity,
or a ``cold front''. A similar structure appears to be observed in
RXJ1720.1+2638 (Mazzotta et al.\ 2000), although the temperature data for
that cluster are less accurate. In Paper I, we have shown that the cool gas
cloud in A3667 moves with a near-sonic velocity in the ambient hot gas.
While the speed of the cloud in A2142 cannot be measured accurately, its
value is consistent with velocities up to the velocity of sound.

A strong suppression of the transport processes, most probably by magnetic
fields, is required to prevent the observed sharp density and temperature
gradients from dissipating. Ettori \& Fabin (2000) argue that the sharp
change of gas temperature in A2142 requires a very significant suppression
of heat conduction relative to the classical value. In A3667, the
suppression of diffusion is observed directly, because the width of the
density discontinuity is smaller than the Coulomb mean free path (Paper I).

In both A3667 and A2142, the cold front appears to be dynamically stable
because it has a smooth shape and shows no visible perturbations within a
certain sector in the apparent direction of the cloud motion. In this Paper,
we use this fact to infer the strength of the intracluster magnetic field
that could naturally provide surface tension stabilizing the front against
the development of the Kelvin-Helmholtz instability.

We use $H_0=50$~km~s$^{-1}$~Mpc$^{-1}$. The derived magnetic field strength
scales as $h^{1/4}$, while the derived ratio of magnetic and thermal
pressures does not depend on $H_0$.

\section{Overview of \chandraintitle\ results}
\label{sec:observations}

The full account of the \emph{Chandra} observation of A3667 is given in
Paper~I, and here we briefly review the relevant details. The most prominent
feature in the \emph{Chandra} image is the sharp 0.5~Mpc-long discontinuity
in the surface brightness distribution located approximately 550~kpc
South-East of the cluster center (Fig.~\ref{fig:edge}). The gas temperature
distribution shows unambiguously that this structure is a boundary between
the large dense body of cool ($T_c=4.1\pm0.2$~keV, 68\% confidence) gas
moving through the hotter ($T_h=7.7\pm0.8$~keV) ambient gas. We call this
structure the ``cold front''. The front shape in projection is very
accurately described by a circle with $R=410\pm15$~kpc.

The gas pressure profile across the front shows that the cloud moves at a
speed close to the velocity of sound in the ambient hot gas (with the Mach
number $M=1\pm0.2$). The measured high velocity of the cloud is
substantiated by the observed compression of the ambient gas near the front,
and a possible bow shock seen in the \emph{Chandra} image further ahead of
the front. This moving cloud probably is a remnant of one of the large
subclusters that have merged recently to form the A3667 cluster.

\begin{inlinefigure}
  \medskip
  \centerline{\includegraphics{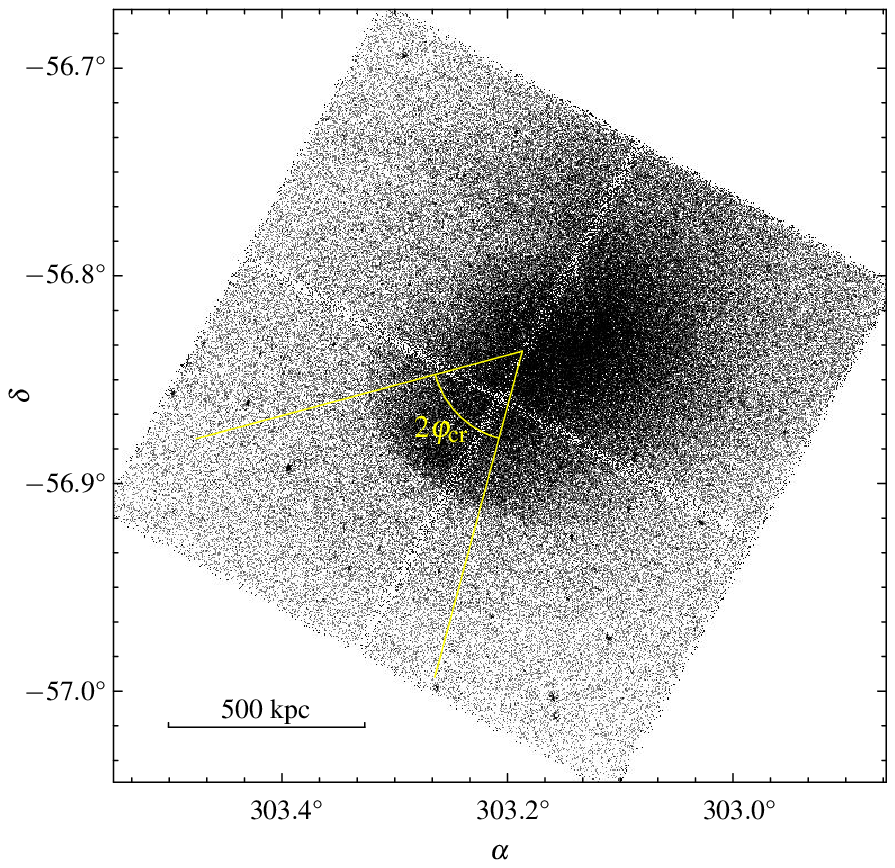}}
  \caption{\emph{Chandra} 0.5--4~keV image of A3667. The surface brightness
    edge (cold front) is sharp within the sector $2\varphi_{\rm cr}=60\degree$,
    outside which it gradually disappears.\label{fig:edge}}
\end{inlinefigure}

\section{Magnetic field and the structure of the front}

\begin{figure*}
  \centerline{\includegraphics{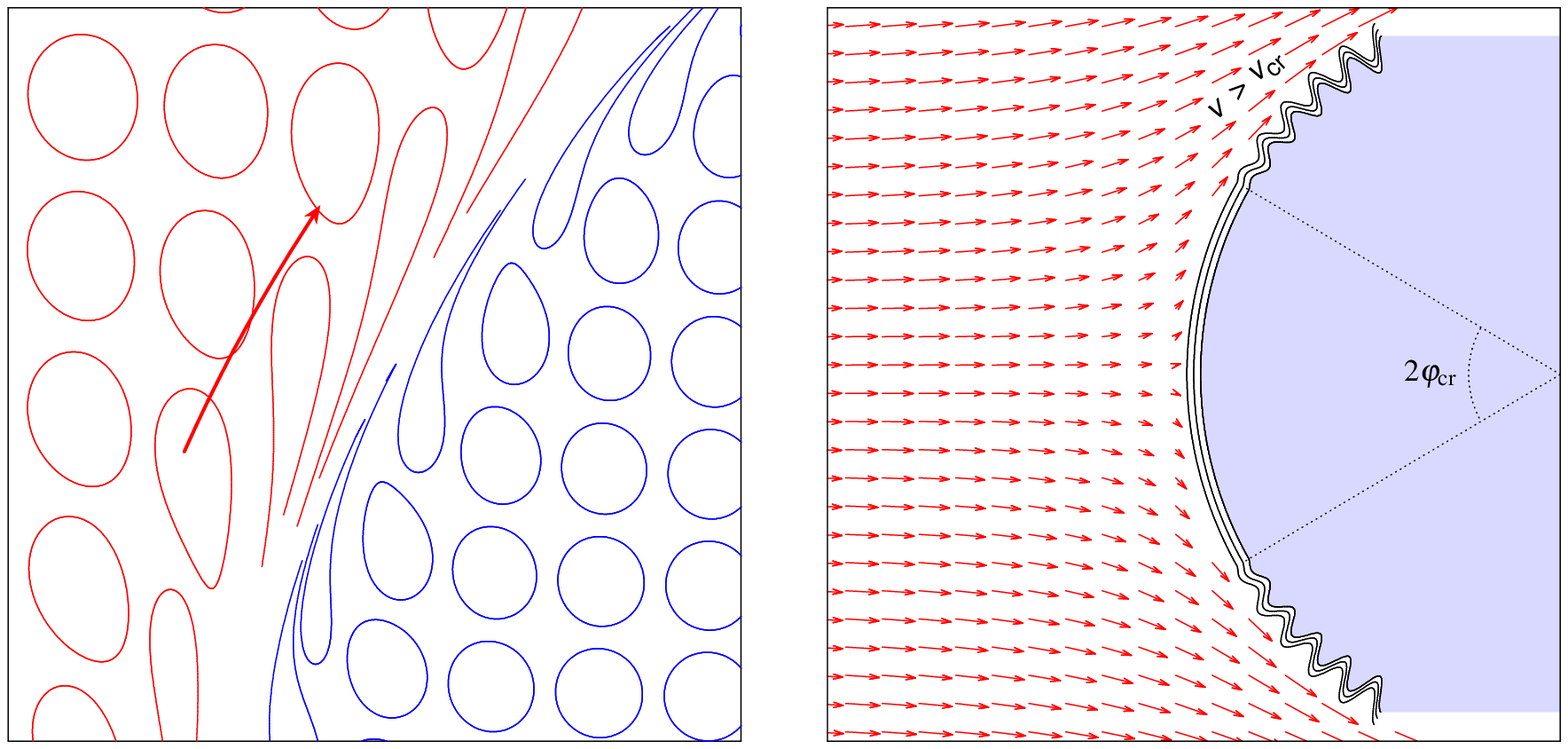}}
  \caption{\emph{(a)} Illustration of the formation of the magnetic
    layer near the front surface. The initially tangled magnetic lines in
    the ambient hot gas (red) are stretched along the surface because of
    tangential motion of the gas. The magnetic lines inside the front are
    stretched because, in the absence of complete magnetic isolation, the
    cool gas experiences stripping. This process can form a narrow layer in
    which the magnetic field is parallel to the front surface. Such a layer
    would stop the transport processes across the front, as well as further
    stripping of the cool gas. \emph{(b)} The interface between the cool and
    hot gas is subject to the Kelvin-Helmholtz instability. The magnetic
    layer can suppress this instability in the region where the tangential
    velocity is smaller that a critical value $V_{\rm cr}$. The velocity
    field shown for illustration corresponds to the flow of incompressible
    fluid around a sphere.}\label{fig:mag}
\end{figure*}

The observed width of the front is very small. Even with the \emph{Chandra}
resolution, the surface brightness profile across the front is consistent
with projection of an infinitely narrow gas density discontinuity. The upper
limit on its width, $3.5''$ or $5$~kpc, is 2--3 times smaller than the
Coulomb mean free path (Paper I).  This requires suppression of the
diffusion by a magnetic field. A magnetic configuration which can stop the
diffusion should indeed arise in a narrow boundary region between the two
tangentially moving gas layers, through a process schematically shown in
Fig.~\ref{fig:mag}. As the ambient gas flows around the dense cloud and the
cool gas is stripped from the surface of the cloud, magnetic field lines,
which are frozen into the gas and initially tangled, should stretch along
the front. When a layer with the parallel magnetic field forms, it would
prevent further stripping of the cool gas and also stop microscopic
transport processes across the boundary, explaining the observed small width
of the front.

Such a layer with the parallel magnetic field also would support the
dynamical stability of the front. At some distance from the leading edge of
the cloud, the gas exterior to the front should develop a high tangential
velocity. The interface between the tangentially moving gas layers normally
is unstable due to the Kelvin-Helmholtz (\KH) instability. We show below
that in the absence of any stabilizing factors, the \KH\ instability should
disturb the front surface outside $\sim10\degree$ of the direction of the cloud
motion. Figure~\ref{fig:edge} shows that the front has a smooth circular
shape and remains sharp within the sector $\varphi_{\rm cr}\approx30\degree$ in
the direction of motion, and hence should be dynamically stable. We suggest
that the \KH\ instability is suppressed by surface tension of the frozen-in
magnetic field, provided that it is parallel to the front. At larger angles
($\varphi>30\degree$) from the direction of motion, the front becomes less
sharp and gradually vanishes. We interpret this as the onset of the \KH\ 
instability, when the tangential velocity of the ambient gas exceeds a
critical value beyond which the instability cannot be suppressed by the
magnetic field (cf.\ Fig.~\ref{fig:mag}b). Assuming that the interface
between two gases is barely unstable at $\varphi=30\degree$, we can estimate
the required magnetic field strength.

Our further discussion is organized as follows. In
\S~\ref{sec:tang:velocity}, we estimate the distribution of the tangential
velocity of the ambient gas near the leading edge of the cloud.  In
\S~\ref{sec:kh:hydro}, we demonstrate that if the Kelvin-Helmholtz
instability is not suppressed, its growth time is smaller than the cluster
core passage time for angles $\varphi>10\degree$, and therefore the instability
already should have disturbed the front. In \S~\ref{sec:kh:suppression}, we
estimate the magnetic field strength from the condition that the
Kelvin-Helmholtz instability is suppressed within the sector
$\varphi<\varphi_{\rm cr}$, but develops at larger angles.

For simplicity, we assume that the front moves through the ambient gas with
the Mach number $M=1$ exactly (while the measured value is $M=1\pm0.2$).

\section{Tangential Velocity Along the Front Surface}
\label{sec:tang:velocity}

For our purposes, the 3D shape of the cloud can be approximated by a
cylinder with a $\sim 250$~kpc radius that has a rounded head with the
curvature radius $R=410$~kpc seen as a front in projection. The radius of
the region of interest (an area on the front within $\varphi\lesssim30\degree$)
is smaller than the radius of the cylinder.  Therefore, we can adequately
approximate the gas flow in this region by that around a ($R=410$~kpc)
sphere, for which there are published laboratory measurements and numerical
simulations for a range of Mach numbers of the inflowing gas.

It is instructive to consider first the analytic solution for the flow of an
incompressible fluid (e.g., Lamb 1945, see also Fig.~\ref{fig:mag}b). The
inflowing gas slows down near the stagnation point at the leading edge of
the sphere but then reaccelerates to high velocities as it is squeezed to
the sides by new portions of the inflowing gas. The velocity at the surface
of the sphere has only a tangential component,
$V_t=\slantfrac{3}{2}\,V\inf\sin\varphi$, where $V\inf$ is the velocity of
the inflowing gas at infinity. Already at $\varphi=30\degree$, the
incompressible gas has $V_t=\slantfrac{3}{4}\,V\inf$.

There is no known analytic solutions for the flow of compressible gas around
a sphere at $M\approx1$, but the qualitative picture is similar. We use the
results of numerical simulations of the $M\inf=1.05$ flow by Rizzi (1980),
whose results are in good agreement with the published laboratory
measurements at slightly higher $M\inf$ (e.g., Gooderman \& Wood 1950).  The
velocity field of a compressible gas flow is parameterized by the spatial
distribution of the local Mach number. The distribution of the Mach number
on the surface of the sphere derived from Rizzi's simulation can be
approximated as
\begin{equation}\label{eq:rizzi:m}
  M\simeq1.1\sin\varphi.
\end{equation}
At $\varphi=30\degree$, the local Mach number is $M=0.55$. In absolute units,
the gas velocity at this point is $V=0.61\,V\inf$, i.e.\ only 20\% lower
than that in the incompressible solution.

\section{Kelvin-Helmholtz Instability Without Magnetic Field}
\label{sec:kh:hydro}

Consider the flat interface between cold gas at rest and a hotter gas
flowing at the local Mach number $M$. The velocity of the flow can be
sufficiently large so that the compressibility of the gas is important. In
the absence of stabilizing factors such as gravity or surface tension, the
interface is unstable due to the Kelvin-Helmholtz instability (see, e.g.,
Gerwin 1968 for a review).  The wave vector of the fastest-growing mode is
in the direction of the flow. The growth time of this mode can be found from
the dispersion equation (cf.\ eq.~5.5 in Miles 1958):
\begin{equation}\label{eq:dispers}
  -\frac{1}{\omega^2}-\frac{c_h^2/c_c^2}{(\omega-Mc_hk)^2}+
  \frac{1}{k^2c_c^2}=0,
\end{equation}
where $c_c$ and $c_h$ are the velocities of sound in the cold and hot gas,
respectively, and $k$ and $\omega$ is the wave number and frequency. In
solving this equation, we should take into account the adiabatic compression
of the hot gas as it arrives at the surface of the cold cloud from infinity,
where it flows at our measured Mach number $M\inf\approx1$. The velocity of
sound as a function of the local Mach number is (e.g., Landau \& Lifshitz
1959)
\begin{equation}
  \frac{c_h^2}{c\inf^2}=\frac{T_h}{T\inf}=
  \frac{1+\slantfrac{1}{2}\,(\gamma-1)M\inf^2}
  {1+\slantfrac{1}{2}\,(\gamma-1)M^2},
\end{equation}
where $c\inf$ is the velocity of sound in the inflowing gas at infinity,
$c\inf=(1.37\pm0.07)c_c$ for the actual temperatures inside and outside our
cloud (\S~\ref{sec:observations}), and $\gamma=\slantfrac{5}{3}$ is the
adiabatic index. For plausible values of $M$, equation~(\ref{eq:dispers})
has two complex roots, one of which corresponds to the growing mode. The
instability growth time is $\tau=(\Im\,\omega)^{-1}$. Perturbations of all
scales are unstable, but for some scales, the instability may develop slowly
compared to the lifetime of the front. The relevant time scale with which
$\tau$ should be compared is the cluster core passage time, $t_{\rm
  cross}=L/M\inf c\inf$, where $L$ is the cluster size. The value of
$\exp(t_{\rm cross}/\tau)$ is the perturbation growth factor over the period
of time it takes the cool gas cloud to travel the distance $L$; if $t_{\rm
  cross}/\tau>1$, the perturbation is effectively unstable.

For the observed gas temperatures and $M\inf=1$, the numerical solution of
eq.~(\ref{eq:dispers}) for $M<1$, the regime relevant for our region of
interest, gives the approximate relation $t_{\rm cross}/\tau=0.075M\, L/l$,
where $l=2\pi/k$ is the perturbation wavelength. Combining this with the
distribution of the Mach number on the surface of the sphere
(eq.~\ref{eq:rizzi:m}), we have
\begin{equation}
  \frac{t_{\rm cross}}{\tau} \simeq 0.083\, \frac{L}{l}\, \sin\varphi.
\end{equation}
For a cluster-scale length of $L=1$~Mpc, we find that a 10~kpc perturbation
becomes effectively unstable already for $\varphi=7\degree$. At
$\varphi=30\degree$, perturbations with $l<45$~kpc are unstable. However, the
front appears undisturbed within at least 30\degree\ around the direction of
motion. Recall that the observed width of the front is smaller than 5~kpc,
and a 45~kpc widening would be easily observable. Therefore, the \KH\ 
instability of the cold front must be suppressed.

Long-wavelength \KH\ perturbations can be stabilized by a gravitational
field perpendicular to the interface (e.g., Lamb 1945). However, we find
that for the gravitational acceleration estimated from the gas pressure
gradient outside the cloud, only $l>4500 M^2$~kpc modes are stable, and
therefore gravity is unimportant for the wavelengths of interest.

\section{Suppression of the Kelvin-Helmholtz Instability by Magnetic Field}
\label{sec:kh:suppression}

The \KH\ instability can be suppressed by the surface tension of a magnetic
field, if the magnetic field is parallel to the interface and to the
direction of the flow. Our considerations can be greatly simplified by the
fact that the gas can be treated as incompressible on both sides of the cold
front in the region of interest, $\varphi\lesssim30\degree$.  The outer gas can
be considered incompressible because it flows with a relatively low local
Mach number $M\lesssim0.5$. The cool gas can be considered incompressible
because the growing modes of the \KH\ instability generally have low phase
speed.  Indeed, the solution of eq.~(\ref{eq:dispers}) obtained above is
very close to the solution in the incompressible limit $c_h,c_c\rightarrow
\infty$ for the gas velocities of interest. The interface between the two
incompressible magnetized fluids is stable if
\begin{equation}\label{eq:kh:mag}
  B_h^2+B_c^2 > 4\pi\, \frac{\rho_h\,\rho_c}{\rho_h+\rho_c}\, V^2,
\end{equation}
where $B_h$ and $B_c$ is the magnetic field strength in the hot and cool
gas, $\rho_h$ and $\rho_c$ are the gas densities, and $V$ is tangential
velocity of the hot gas (e.g., Landau \& Lifshitz 1960, \S~53).
If the magnetic pressure is small compared to the thermal pressure, $p$ (as
it turns out to be the case), or is the same fraction of $p$ on both sides
of the interface, the stability condition~(\ref{eq:kh:mag}) can be rewritten
in terms of the thermal pressure and temperature of the gas:
\begin{equation}
  \frac{B_h^2}{8\pi}+\frac{B_c^2}{8\pi}>\frac{1}{2}\,\frac{\gamma
    M^2}{1+T_c/T_h}  p.
\end{equation}
The stability of the cold front within $\varphi<30\degree$ of the direction of
the cloud motion, where $M\le0.55$, requires $(B_h^2+B_c^2)/8\pi>0.17p$ for
the observed ratio of $T_c/T_h$. If smearing of the surface brightness edge
at angles $\varphi\gtrsim30\degree$ is interpreted as the development of the \KH\
instability, this lower limit becomes a measurement of the magnetic
pressure.

If magnetic field is too weak to suppress the \KH\ instabilty completely, it
still increases the instability growth time, which is then
\begin{equation}
  \label{eq:kh:mag:growth}
  (\Im\,\omega)^{-1} =k^{-1}\left[V^2 \frac{\rho_c\rho_h}{(\rho_c+\rho_h)^2} - \frac{B_h^2+B_c^2}{4\pi(\rho_c+\rho_h)}\right]^{-1/2}.
\end{equation}
Thus, even with a weaker magnetic field, the interface can be effectively
stable on the time scale of the cluster core passage. However,
eq.~(\ref{eq:kh:mag:growth}) shows that this quickly becomes unimportant as
the magnetic field strength decreases below the stability limit. For
example, for $B$ of only 0.7--0.8 of the stability limit, the growth time is
1.1--1.4 of its value for $B=0$.

The measurement uncertainty in the derived value of the magnetic field
strength is mostly due to uncertainty in the angle, $\varphi_{\rm cr}$,
where the interface becomes unstable, and hence in the local Mach number of
the hot gas. For $\varphi_{\rm cr}$ in the range $30\degree\pm10\degree$ (which
appears to be a conservative interval, see Fig.~\ref{fig:edge}), we find
from eq.~(\ref{eq:kh:mag}) and (\ref{eq:rizzi:m}) that the magnetic field
strengths are in the interval
\begin{equation}\label{eq:pmag:contsraint}
 0.09p<(B_h^2+B_c^2)/8\pi<0.23p.
\end{equation}

Our method directly constrains only the sum of magnetic pressures on the two
sides of the interface. The maximum of the magnetic field strengths in the
cold and hot gases, $B=\max(B_h,B_c)$, is in the interval $(4\pi p_{\rm
  mag})^{1/2}<B<(8\pi p_{\rm mag})^{1/2}$, where $p_{\rm mag} =
(B_h^2+B_c^2)/8\pi$ is constrained by (\ref{eq:pmag:contsraint}). Using the
value of gas pressure inside the dense cloud derived in Paper~I, we find the
corresponding uncertainty interval $7\,\mbox{\muG}<B<16\,\mbox{\muG}$.

\section{Discussion}

We showed that the sharpness of the front at small angles from the direction
of the cloud motion and its smearing at larger angles can be explained by
the existence of a layer in which the magnetic field is roughly parallel to
the front and has a strength of order $10\,$\muG. As we discussed above,
such a layer may form by stretching the magnetic field lines near the front
by tangential gas motions. As a result, the magnetic field strength in this
layer is probably higher than in the rest of the intracluster gas. The
magnetic field amplification on the surface of the plasma cloud is
discussed, for example, in Jones, Ryu \& Tregillis (1996). An important
conclusion is that the magnetic pressure is only a small fraction of the
thermal pressure: $p_m/p\sim0.1-0.2$ in the magnetic layer and still lower
in the rest of the cluster. If the magnetic pressure were of order of the
thermal pressure, the front would be stable for Mach numbers up to
$M\sim2(1+T_c/T_h)^{1/2}\gamma^{-1/2}=1.9$, i.e.\ over the entire surface of
the cool gas cloud.

Faraday rotation measurements of the magnetic field outside the cluster
cooling flows usually give the values of $\sim 1$~\muG\ (Kim et al.\ 1991).
Since Faraday rotation gives the integral of the magnetic field along the
line of sight, there is a considerable uncertainty of the absolute field
strength due to unknown degree of entanglement. Near the cold front, the
field is straightened and amplified, so our measurement provides a likely
upper limit on the absolute field strength in the cluster core.

Finally, we note that our arguments would become invalid if the shape of the
front were non-stationary, for example, if the cool gas cloud expands into
the low-density medium (as opposed to moving as a whole). However, the front
velocity relative to the ambient gas exceeds the velocity of sound within
the cloud. Since the \emph{in situ} acceleration of a gas cloud to the
supersonic velocity is impossible, the cool gas cloud must have arrived from
outside the main cluster. In this case, the fact that the shape of the front
remains smooth and circular after traveling a large distance, suggests that
the front is in fact close to stationary, and therefore our arguments
regarding the magnetic field are valid.

\acknowledgments

We thank P.~Nulsen, W.~Forman, and C.~Jones for helpful comments.  This
study was supported by NASA grant NAG5-9217 and contract NAS8-39073.


\begin{references}

  \reference{} Ettori, S., \& Fabian, A. C. 2000, MNRAS, in press
  (astro-ph/0007397)

  \reference{} Fabian, A. C. 1994, ARA\&A, 32, 277

  \reference{} Fusco-Femiano, R.\ et al. 2000, ApJ, 534, L7

  \reference{} Gerwin, R. A. 1968, Rev Mod Phys, 40, 652

  \reference{} Gooderum, P. B. \& Wood, G. P. 1950, NACA Technical Note 2173
  
  \reference{}  Jones, T.~W., Ryu, D., \& Tregillis, I.~L. 1996, ApJ, 473,
  365

  \reference{} Kim, K.~., Kronberg, P.~P., \& Tribble, P.~C. 1991, ApJ, 379,
  80

  \reference{} Lamb, H. 1945, Hydrodynamics (New York: Dover)

  \reference{} Landau, L.~D., \& Lifshitz, E.\ M.\ 1959, Fluid Mechanics
  (London: Pergamon)

  \reference{} Landau, L.~D., \& Lifshitz, E.\ M.\ 1960, Electrodynamics of
  Continuous Media  (New York: Pergamon)

  \reference{} Loeb, A. \& Mao, S. 1994, ApJ, 435, L109

  \reference{} Mazzotta, P. et al.\ 2000, in preparation

  \reference{} Miles, J. W. 1958, J. Fluid Mech., 4, 538

  \reference{} Miralda-Escud\'e, J. \& Babul, A. 1995, ApJ, 449, 18


  \reference{} Rizzi, A. 1980, in Numerical Methods in Applied Fluid
  Dynamics, ed. B. Hunt (London: Academic Press), p~555

  \reference{} Vikhlinin, A., Markevith, M.\ \& Murray 2000, ApJ, submitted
  (astro-ph/0008496)
\end{references}
\end{document}